\documentclass[aps,prb,twocolumn,superscriptaddress,showpacs]{revtex4-1}

\usepackage{amsmath}
\usepackage{graphicx}
\usepackage{calc}
\usepackage{bm}

\usepackage[T2A]{fontenc}
\usepackage[cp1251]{inputenc}
\usepackage[english]{babel}

\begin{document}

\title{Switching ferroelectricity in SnSe across diffusionless martensitic phase transition}

\author{N.N. Orlova}
\author{A.V. Timonina}
\author{N.N. Kolesnikov}
\author{E.V. Deviatov}

\affiliation{Institute of Solid State Physics of the Russian Academy of Sciences, Chernogolovka, Moscow District, 2 Academician Ossipyan str., 142432 Russia}

\date{\today}

\begin{abstract}
We experimentally investigate transport properties of a hybrid structure, which consists of a thin single crystal SnSe flake on a top of 5~$\mu$m spaced Au  leads. The structure initially is in highly-conductive state, while it can be switched to low-conductive one at high currents due to the Joule heating of the sample, which  should be identified as $\alpha$-$Pnma$ -- $\beta$-$Cmcm$ diffusionless martensitic phase transition in SnSe. For highly-conductive state, there is significant hysteresis in $dI/dV(V)$ curves at low biases,  so the sample conductance depends on the sign of the applied bias change. This hysteretic behavior reflects slow relaxation due to additional polarization current in the  ferroelectric SnSe phase, which we confirm by direct measurement of time-dependent relaxation curves.  In contrast, we observe no noticeable relaxation or low-bias hysteresis for the quenched $\beta$-$Cmcm$ low-conductive phase. Thus, ferroelectric behavior can be switched on or  off in transport through hybrid SnSe structure by controllable $\alpha$-$Pnma$ -- $\beta$-$Cmcm$ phase transition. This result can also be  important for nonvolatile memory development, e.g.  phase change memory for neuromorphic computations  or other applications in artificial intelligence and modern electronics.
\end{abstract}

\pacs{71.30.+h, 72.15.Rn, 73.43.Nq}

\maketitle

\section{Introduction}

 Recent interest to chalcogenides is connected with their electronic, thermal and optical properties~\cite{progress MX}.  For pure science, topological semimetals~\cite{armitage} can be realized as three-dimensional single chalcogenides crystals. Also, they are usually considered as a platform for creating  modern two-dimensional materials,  due to the layered structure and  high carrier mobility.~\cite{synthesis MX} For applications, they are mostly regarded for  photodetectors, phototransistors or different photovoltaic devices~\cite{Phdet,PhL,Op,Pht1,Pht2}. They also demonstrate thermoelectricity~\cite{TE1,TE2,TE3}, piezoelectricity~\cite{PE1,PE2} and ferroelectricity~\cite{FESnSe,FESnTe,FESnS} even at room temperature. 

Ferroelectric properties of different chalcogenides are of great interest due to the underlying physics and potential applications~\cite{FEreview}. Recently, three-dimensional WTe$_2$ single crystals were found to demonstrate  coexistence of metallic conductivity and ferroelectricity at room temperature~\cite{WTe2_fer}. Out-of-plane spontaneous polarization of ferroelectric domains is found to be bistable, it can be affected by high external electric field. Also, in-plane ferroelectric polarization has been also demonstrated~\cite{FESnSe} for layered SnSe monochalcogenide~\cite{SnSeprop}.  
 
Ferroelectric properties are unambiguously connected with crystal symmetry~\cite{WTe2_fer,ferr_book}. Monochalcogenides are often centrosymmetric in the bulk but lack inversion symmetry~\cite{mendoza} in single-layer~\cite{FESnTe,FESnS1} or few-layer~\cite{FESnTe-few,FESnS-few} forms.  On the other hand, crystal structure is subjected to modification in monochalcogenides across phase transition. For example, layered SnSe exhibits a low degree of lattice symmetry, with a distorted NaCl structure and an in-plane anisotropy~\cite{SnSeprop}. It belongs to orthorhombic $\alpha$-$Pnma$ phase at room temperature, which can be converted~\cite{PhTSnSe_SnS}  into the symmetric  $\beta$-$Cmcm$ phase at moderate temperatures. This transition should seriously affect, in particular, spontaneous ferroelectric polarization. Thus, correlation between the ferroelectric properties and crystal symmetry can be experimentally studied in chalcogenides, being also  important for nonvolatile memory investigations~\cite{PhTbylaser}, e.g. for phase change memory for neuromorphic computations  or other applications in artificial intelligence and modern electronics.

 Here, we experimentally investigate transport properties of a hybrid structure, which consists of a thin single crystal SnSe flake on a top of 5~$\mu$m spaced Au  leads. The structure initially is in highly-conductive state, while it can be switched to low-conductive one at high currents due to the Joule heating of the sample, which  should be identified as $\alpha$-$Pnma$ -- $\beta$-$Cmcm$ phase transition in SnSe. For highly-conductive state, there is significant hysteresis in $dI/dV(V)$ curves at low biases,  so the sample conductance depends on the sign of the applied bias change. This hysteretic behavior reflects slow relaxation due to additional polarization current in the  ferroelectric SnSe phase, which we confirm by direct measurement of time-dependent relaxation curves.  In contrast, we observe no noticeable relaxation or low-bias hysteresis for the quenched $\beta$-$Cmcm$ low-conductive phase. Thus, ferroelectric behavior can be switched on or  off in transport through hybrid SnSe structure by controllable $\alpha$-$Pnma$ -- $\beta$-$Cmcm$ phase transition. This result can also be  important for nonvolatile memory development.

\section{Samples and techniques}

\begin{figure}[t]
\center{\includegraphics[width=\columnwidth]{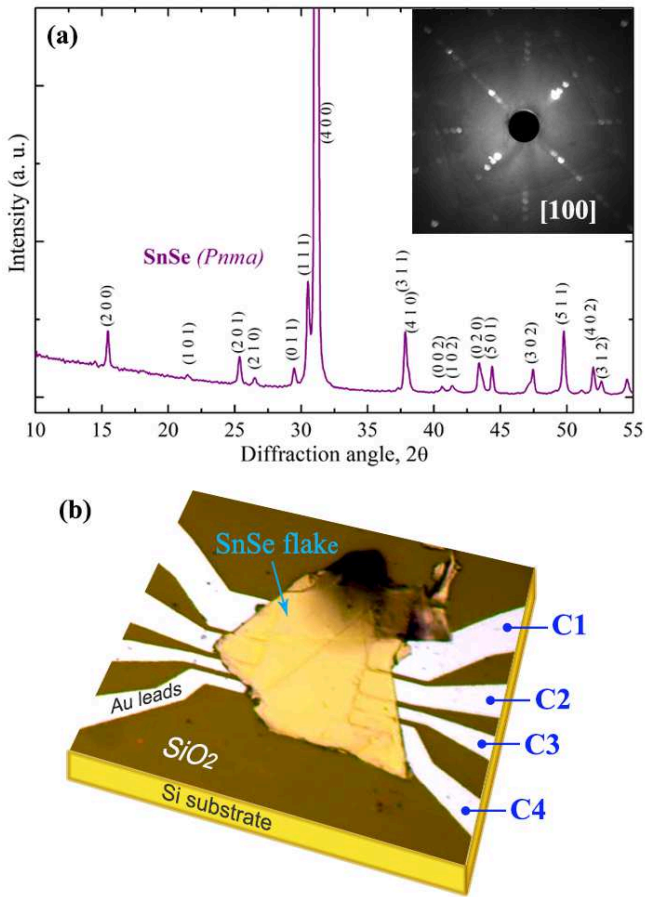}}
\caption{ (a)  X-ray diffraction pattern, which confirms single-phase SnSe with orthorhombic crystal system ($Pnma$(62) space group, $a=11.502(6)$~\AA, $b= 4.153$~\AA, and $c=4.450$~\AA). Inset shows the Laue x-ray diffraction pattern of SnSe single crystal, the  x-ray direction is perpendicular to the exfoliated planes with zone axis [100]. (b) Optical image of a sample. SnSe flake is  placed over 100~nm thick Au leads, the leads profile can be seen under the ultra-thin~\cite{SnSeprop} (below 100~nm) flake. The relevant (bottom) SnSe surface is protected from any oxidation or contamination by SiO$_2$ substrate.   The leads C1, C2, C3, C4 are separated by 5~$\mu$m intervals, they are used for two- and four-point electrical  measurements, see the main text.}
\label{xrd sample}
\end{figure}

SnSe compound was synthesized by reaction of selenium vapors with the melt of high-purity tin in evacuated silica ampoules. X-ray diffraction pattern is shown in Fig.~\ref{xrd sample} (a). It confirms single-phase SnSe with orthorhombic crystal system ($Pnma$(62) space group, $a=11.502(6)$~\AA, $b= 4.153$~\AA, and $c=4.450$~\AA). The SnSe single crystals with diameter 20 mm and length 80 mm were grown by vertical zone melting in silica crucibles under argon pressure.
The inset to Fig.~\ref{xrd sample} (a) demonstrates the Laue x-ray diffraction pattern of SnSe single crystal, the x-ray direction  is perpendicular to the exfoliated planes with zone axis [100].

In layered monochalcogenides, ferroelectricity appears below some critical thickness~\cite{FEreview}, which can be estimated~\cite{SnSeprop} as 200~nm for SnSe. 
For electrical measurements, an ultra-thin~\cite{SnSeprop} SnSe flake (below 100~nm) is placed on the top of the pre-defined Au leads pattern~\cite{II-WSM3,II-WSM4,cdas,timnal}, see Fig.~\ref{xrd sample} (b). This preparation procedure allows to specify experimental geometry for charge transport, and, simultaneously, it  minimizes chemical or thermal treatment of the initial flake. The relevant (bottom) SnSe surface is also protected from any oxidation or contamination by SiO$_2$ substrate.  

Standard photolithography and lift-off technique are used to define 100~nm thick Au leads on the insulating SiO$_2$ substrate. Thin SnSe  flakes are obtained from the initial ingot by regular mechanical exfoliation, also known as scotch-tape technique. Afterwards, the exfoliated flake is transferred on the top of the leads pattern, as it is shown in Fig.~\ref{xrd sample} (c). The flake is slightly pressed to the leads by another oxidized silicon substrate. Weak pressure is applied with a special metallic frame, which keeps the substrates strictly parallel. This procedure  has been verified  to provide electrically stable contacts with highly transparent  metal-semiconductor interfaces~\cite{II-WSM3,II-WSM4,cdas,timnal}.

The sample resistance is about 10--100~kOhm, the actual value depends on a particular Au-SnSe interface, overlap area and SnSe flake thickness.

 For correct measurement of high resistances (about 100~kOhm), one have to directly apply voltage bias $V$ to one of the contacts in Fig.~\ref{xrd sample} (c) in respect to the neighbor (grounded) lead, current $I$ is measured in the circuit. To obtain differential conductance dependence $dI/dV(V)$, the applied dc bias $V$ is additionally modulated by a small (10~mV) ac component at a frequency of 1100 Hz. The ac current component is measured by lock-in, being proportional to differential conductance $dI/dV$ at a given bias voltage $V$. We verify that the obtained $dI/dV$ value does not depend on the modulation frequency in the range 1~kHz--10~kHz, which is determined by  applied filters.

The samples  are at room temperature initially, since the spontaneous ferroelectric polarization has been demonstrated~\cite{FESnSe} in layered SnSe for   the 
low-symmetry $\alpha$-$Pnma$ phase. At high currents, however, local Joule heating of the sample can be expected up to about 500$^\circ$~C between two neighbor leads, which we have tested by local decomposition of black phosphorus flakes~\cite{black}.

\section{Experimental results}

\begin{figure}[t]
\center{\includegraphics[width=\columnwidth]{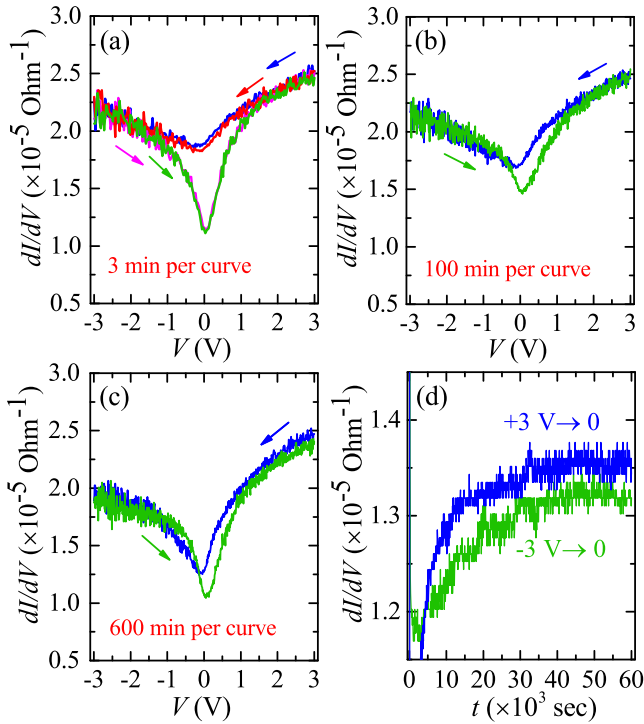}}
\caption{ (a-c) Two-point $dI/dV(V)$ curves for typical 100~kOhm sample, arrows indicate sweep directions for the curves of the same colors. The curves coincide perfectly for the same sweep direction in (a), while there is noticeable hysteresis within the $\pm$1~V range for opposite ones. Also, $dI/dV$ differential conductance is excellently stable at $\pm$3~V, which are the boundaries of the voltage sweep range. The sweep rate is diminishing from (a) to (c): the curves in (a) are obtained for 3 min sweep for the $\pm$3~V range,  it takes 100 min and 600 min for curves in (b) and (c), respectively. The hysteresis amplitude dependence on the sweep rate reflects some slow relaxation process. (d) Time-dependent relaxation  $dI/dV(t)$ at zero bias. The sample is kept at a fixed dwelling voltage (+3~V or -3~V) for 10~min, voltage bias is abruptly set to zero value afterward. $dI/dV(V=0)$ values are  clearly different  for the dwelling voltages of  opposite signs. 
}
\label{IV_3V}
\end{figure}

Fig.~\ref{IV_3V} shows the examples of two-point $dI/dV(V)$ curves for typical 100~kOhm sample. Differential conductance  $dI/dV$ non-linearly increases for positive and negative bias voltages in Fig.~\ref{IV_3V} (a).  In addition, we observe significant hysteresis with bias sweep direction, which somewhat analogous to the previously reported~\cite{wte2mem} for the ferroelectric chalcogenide WTe$_2$: for opposite voltage sweep directions, there is noticeable discrepancy between the curves within the $\pm$1~V range. It is important, that $dI/dV(V)$ curves coincide perfectly in Fig.~\ref{IV_3V} (a) if they are obtained for the same sweep direction, see  red and blue or pink and green curves, respectively. Also, $dI/dV$ differential conductance is excellently stable at $\pm$3~V, which are the boundaries of the voltage sweep range.  Thus, the observed hysteresis can not be ascribed to any artificial drifts, etc., and is defined by the sweep direction.

 The hysteresis amplitude depends on the sweep rate, see Fig.~\ref{IV_3V} (a-c), so it reflects some slow relaxation process.  The latter  can be demonstrated by direct relaxation measurements in Fig.~\ref{IV_3V} (d). To obtain $dI/dV(t)$ curves,   the sample is kept at a fixed dwelling voltage (+3~V or -3~V) for 10~min, voltage bias is abruptly set to zero value afterward.  The time-dependence of $dI/dV(V=0)$ is recorded, which demonstrates slow relaxation for significant time interval. However, $dI/dV(t)$ curves show  clearly different $dI/dV(V=0)$ values for the dwelling voltages of  opposite signs, which confirms finite hysteresis amplitude for the lowest sweep rate in Fig.~\ref{IV_3V} (c).

\begin{figure}[t]
\center{\includegraphics[width=\columnwidth]{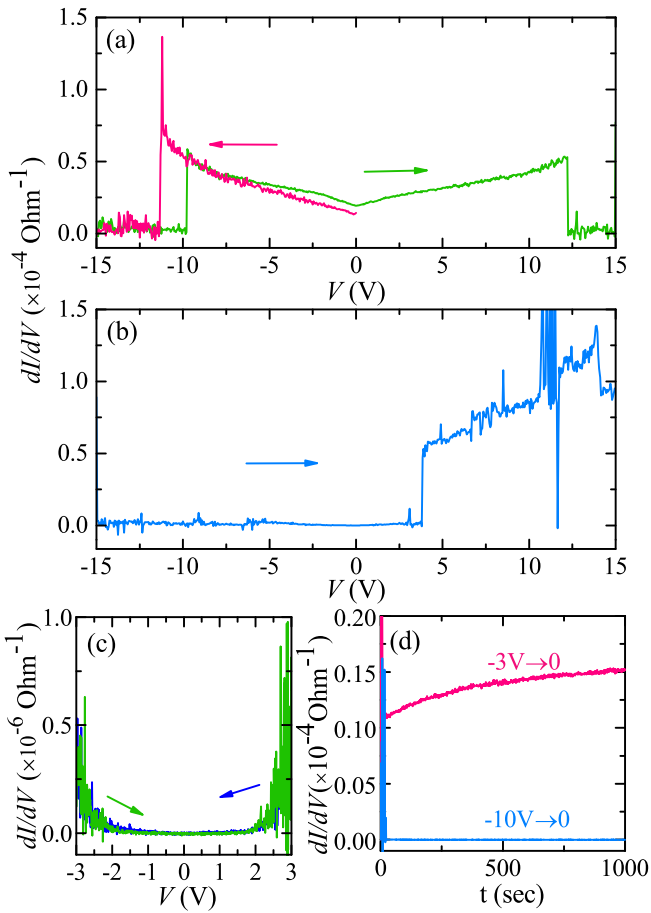}}
\caption{ (a-b) Switching between highly- and low-conductive states of the sample, which we  identify as  $\alpha$-$Pnma$ -- $\beta$-$Cmcm$ phase transition in SnSe. The arrows indicate sweep directions for $dI/dV(V)$ curves. (a) Differential conductance $dI/dV$ falls down in two orders of magnitude at $\approx$-12~V.  $dI/dV$ level is recovered for slow backward bias sweep at $\approx$-10~V, i.e.  with some voltage delay. $dI/dV(V)$ also falls down at $\approx$+12~V, so the threshold voltage  is approximately the same for both bias polarities. (b) The  low-conductive state can be preserved ("quenched") down to  zero bias. (c) There is no hysteresis for the "quenched" low-conductive state, $dI/dV(V)$ curves are  independent of the voltage sweep direction. (d) No relaxation can be observed for the  "quenched" low-conductive state (blue curve), in contrast to the pronounced relaxation from the  highly-conductive one (red curve).}
\label{switching}
\end{figure}

To our surprise, the described above behavior can be controllably switched off or on by applying voltage bias above some threshold value, see   Fig.~\ref{switching}. In a slow voltage sweep from zero to high  negative values in Fig.~\ref{switching} (a) (the red curve), differential conductance $dI/dV(V)$ falls down in two orders of magnitude at approximately $\approx$-12~V. This step-like conductance jump  is found to be reversible, $dI/dV$ level is recovered for slow backward bias sweep, although with some voltage delay (at $\approx$-10~V), see the left part of Fig.~\ref{switching} (a) (the green curve). While continuing sweep from zero to  positive biases, $dI/dV(V)$ also falls down at $\approx$+12~V in the right part of Fig.~\ref{switching} (a), so the threshold voltage  is approximately the same for both bias polarities.  Thus, it seems to be controlled by bias amplitude (e.g., due to local Joule heating) but not the electric field direction. 

For backward sweep from the high biases, the  low-conductive state can either be preserved down to the zero bias value, see Fig.~\ref{switching} (b). In this case, there is no hysteresis in the narrow bias range, see Fig.~\ref{switching} (c). $dI/dV(V)$ curves are flat and independent of the voltage sweep direction, in contrast to ones in  Fig.~\ref{IV_3V}. Also, no relaxation can be observed for the  "quenched" low-conductive state, as demonstrated by the blue curve in Fig.~\ref{switching} (d).  This "quenched" zero-bias state is not stable, it can be switched back to the highly-conductive one by moderate biases, as depicted in the right part of  Fig.~\ref{switching} (b).

We wish to mention the clearly visible noise in experimental curves in Fig.~\ref{IV_3V} (a-c) and  and Fig.~\ref{switching} (c). The noise level is minimal at zero bias, it is obviously increasing with $dI/dV$ signal amplitude. The noise fluctuations are nearly reproducible for the curves, which are recorded in the same sweep direction. This behavior is just opposite to the expected for external electrical noise. It reflects some bias-dependent conductance fluctuations, which can be expected for ultra-thin flakes.

\begin{figure}[t]
\center{\includegraphics[width=\columnwidth]{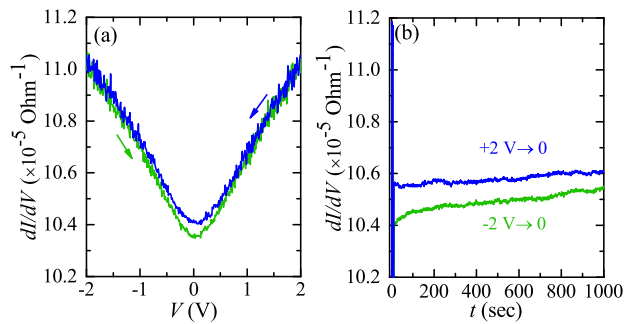}}
\caption{ (a) Two-point $dI/dV(V)$ curves for another SnSe sample with much higher initial zero-bias conductance. There is noticeable hysteresis with the bias sweep direction. (b)  Time-dependent $dI/dV(t)$ relaxation curves at zero bias. All the data are qualitatively similar to ones  from the sample in  Fig.~\ref{IV_3V}. In general, hysteretic $dI/dV(V)$ behavior with slow relaxation is analogous to ferroelectricity-induced one~\cite{wte2mem} for another ferroelectric chalcogenide WTe$_2$.}
\label{IIsample IV}
\end{figure}

Similar $dI/dV(V)$ behavior can be demonstrated for different samples, e.g.  with much higher ($\approx 10$~kOhm) initial  zero-bias conductance in Fig.~\ref{IIsample IV} (a). Two-point  $dI/dV(V)$  increases non-linearly at positive and negative bias voltages, there is noticeable hysteresis with the sweep direction, the noise level is  increasing with $dI/dV$ signal amplitude. The relaxation $dI/dV(t)$ curves  also confirm slow relaxation process with two  different $dI/dV(V=0)$  resistance states for the dwelling voltages of  opposite signs in Fig.~\ref{IIsample IV} (b).

\begin{figure}[t]
\center{\includegraphics[width=\columnwidth]{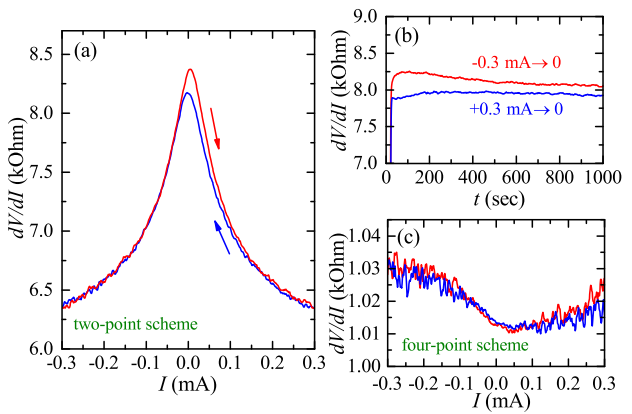}}
\caption{ $dV/dI(I)$ curves for the sample from Fig.~\ref{IIsample IV}, obtained  both in two-point (a) and four-point  (c) techniques. In this case, $I$ is defined and ac voltage drop $\sim dV/dI$ is measured. (a) Two-point  curves are just inverted in respect to ones in Fig.~\ref{IIsample IV} (a), they demonstrate the same hysteretic behavior, which is confirmed by slow $dV/dI(t)$  relaxation  curves in (b). (c)  Four-point curves $dV/dI(I)$ reflect the bulk SnSe differential resistance. They demonstrate no hysteresis, the general $dV/dI(I)$ behavior is obviously different from the two-point one in (a). Thus, slow relaxation in (b) and consequent hysteresis in (a) originate from the Au-SnSe contacts' areas.}
\label{IV other scheme}
\end{figure}

$10$~kOhm samples also allow to directly define current $I$ in the circuit, therefore,  to measure the resulting voltage drop $V$ both in two-point and four-point techniques. For the two-point connection scheme, $dV/dI(I)$ curves of sample  differential resistance are presented in Fig.~\ref{IV other scheme} (a)  for two current sweep directions. These curves are just inverted in respect to ones in Fig.~\ref{IIsample IV} (a), they demonstrate the same hysteretic behavior, which is confirmed by $dV/dI(t)$  relaxation  curves in Fig.~\ref{IIsample IV} (b).

$dV/dI(I)$ behavior is drastically different in the four-point connection scheme, see Fig.~\ref{IV other scheme} (c). The bulk SnSe differential resistance demonstrates no hysteresis, the general $dV/dI(I)$ behavior is obviously different from one in Fig.~\ref{IV other scheme} (a), while the conductance fluctuations are  still increasing with the current bias, maybe, due to the local heating of the bulk SnSe region. Since the contacts are excluded for the four-point measurements, we have to conclude that the results in  Figs.~\ref{IV_3V},~\ref{switching},~\ref{IIsample IV} originate from the Au-SnSe contacts' areas.

\section{Discussion}

As a result, we demonstrate two different, highly- and low-conductive SnSe states. The highly-conductive one exists at low biases, it is characterized by noticeable hysteresis with bias sweep direction. This state can be switched to the low-conductive one, the conductance jump is sharp and can be achieved  for both bias signs. No hysteresis can be observed at low biases in the low-conductive state. 

SnSe is a typical  layered narrow-gap chalcogenide~\cite{SnSeprop,progress MX,indirect}, the initial state is orthorhombic $\alpha$-$Pnma$ phase. In this state, SnSe  is characterized by p-type conductivity~\cite{conduct}, defined by intrinsic vacancies~\cite{vac}. Two-point $dI/dV(V)$ curves reflect the electrochemical potential shift at the Au-SnSe interfaces~\cite{black}, which leads to nearly symmetric $dI/dV$ increase for both bias signs in Figs.~\ref{IV_3V},~\ref{IIsample IV}, and~\ref{IV other scheme} (a). In contrast, there is only negligible $dV/dI(I)$ dependence for the four-point curves in Fig.~\ref{IV other scheme} (b) with excluded Au-SnSe interfaces. 

For SnSe, there is well known martensitic type phase transition~\cite{PhTSnSe_SnS} from initial low-symmetric $\alpha$-$Pnma$ phase to the highly-symmetric  $\beta$-$Cmcm$ one, which  occurs at ~750-800~K\cite{struct}. This temperature can be achieved locally in our setup at high currents,  which we have tested by  decomposition of black phosphorus flakes in similar experimental geometry~\cite{black}. Sharp conductance change is a known fingerprint of martensitic type phase transitions~\cite{Chan}, e.g. it occurs due to the charge density wave formation in layered semiconductors~\cite{Chan,CDW,MnTe}. More generally,  charge density wave formation  accompany any diffusionless phase transition, which is characterized by symmetry change and intrinsic strain. Conductivity falls down at the transition due to the charge density wave pinning by stress, impurities or defects. Also, the high temperature phase can be quenched at room temperature because of the non-thermoelastic character of the transition~\cite{Chan}. 

Thus, we  can identify  $\alpha$-$Pnma$ -- $\beta$-$Cmcm$ phase transition  by differential conductance jump  in Fig.~\ref{switching}. From  Fig.~\ref{switching} (a), the conductance jumps can be observed for both bias polarities, which also confirms Joule heating origin of the effect. In this case,  backward sweep of the bias leads to the quenching of the high-symmetric $\beta$-$Cmcm$ phase down to zero bias. Since the quenched state is not stable, it can be destroyed by moderate heating, as depicted in  Fig.~\ref{switching} (b).

In general, hysteretic $dI/dV(V)$ behavior in Figs.~\ref{IV_3V},~\ref{IIsample IV} and ~\ref{IV other scheme} (a) is qualitatively analogous to ferroelectricity-induced one~\cite{wte2mem} for another ferroelectric chalcogenide WTe$_2$. 

Thin SnSe layers are  characterized~\cite{FESnSe} by in-plane spontaneous ferroelectric polarization at room temperatures.  In this case, any variation of  the source-drain bias leads to the additional polarization current due to the domain wall shift for varying electric field. Indeed, source-drain in-plain field $E_{sd}\sim V/d$ is induced by the voltage bias $V$, where $d=5\mu$m is the separation between the Au leads. The achievable $E_{sd}$ values  ($\sim 10^{5}$~V/m) are too small to align polarization of the whole $SnSe$ flake, so $E_{sd}$ mostly affects the domain wall regions. Any variation of  the domain wall positions lead to the additional polarization current. Since polarization current is connected with lattice deformation in ferroelectrics, we observe it as slow relaxation in $dI/dV$. In other words, the dc circuit of our experimental setup is equivalent to so called Sawyer-Tower's circuit~\cite{ferr_book,scheme}, so the difference between every two curves in  Figs.~\ref{IV_3V},~\ref{IIsample IV} and ~\ref{IV other scheme} (a) represents a standard ferroelectric hysteresis loop. 

In contrast to the bulk WTe$_2$ samples with broken inversion symmetry~\cite{WTe2_fer}, ferroelectricity in SnSe is connected a low degree of lattice symmetry in thin layers~\cite{FEreview}. The critical thickness  can be estimated~\cite{SnSeprop} as 200~nm for SnSe. We observe slow relaxation for ultra-thin~\cite{SnSeprop} flakes in planar experimental geometry, where electric field is mostly concentrated at the Au-SnSe interface. This is the reason not to observe  $dI/dV(V)$ hysteresis in the four-point connection scheme for bulk SnSe resistance, see Fig.~\ref{IV other scheme} (b). Also, polarization current is some addition to the ordinary one for the conductive samples, so the ferroelectric hysteresis loop is more pronounced for thinner (with higher resistance)  samples, cp. in Figs.~\ref{IV_3V} and~\ref{IIsample IV}.   

Ferroelectric properties are unambiguously connected with crystal symmetry~\cite{WTe2_fer,ferr_book}, which can be utilized to confirm ferroelectric origin of the low-bias hysteresis. For example, gold adatom absorption at defects~\cite{NVRS1} can not be sensitive to the crystal symmetry. We do not observe neither low-bias hysteresis (i.e. conductance dependence on the previously applied bias sign) nor slow relaxation  in the quenched highly-symmetric $\beta$-$Cmcm$ phase in  Fig.~\ref{switching} (c) and (d).  Thus, relaxation can be switched on or off by controllable $\alpha$-$Pnma$ -- $\beta$-$Cmcm$ phase transition in SnSe,  which also confirms its ferroelectric origin. This result can also be  important for nonvolatile memory development, e.g.  phase change memory for neuromorphic computations  or other applications in artificial intelligence and modern electronics.

\section{Conclusion}

As a conclusion, we experimentally investigate transport properties of a hybrid structure, which consists of a thin single crystal SnSe flake on a top of 5~$\mu$m spaced Au  leads. The structure initially is in highly-conductive state, while it can be switched to low-conductive one at high currents due to the Joule heating of the sample, which  should be identified as $\alpha$-$Pnma$ -- $\beta$-$Cmcm$ phase transition in SnSe. For highly-conductive state, there is significant hysteresis in $dI/dV(V)$ curves at low biases,  so the sample conductance depends on the sign of the applied bias change. This hysteretic behavior reflects slow relaxation due to additional polarization current in the  ferroelectric SnSe phase, which we confirm by direct measurement of time-dependent relaxation curves.  In contrast, we observe no noticeable relaxation or low-bias hysteresis for the quenched $\beta$-$Cmcm$ low-conductive phase. Thus, ferroelectric behavior can be switched on or  off in transport through hybrid SnSe structure by controllable $\alpha$-$Pnma$ -- $\beta$-$Cmcm$ phase transition. This result can also be  important for nonvolatile memory development.

\section{Acknowledgement}

The authors are grateful to S.S. Khasanov for fruitful discussions, XPS, and x-ray sample characterization. We gratefully acknowledge financial support  by  RF State task.

\end{document}